\documentclass[preprint,12pt]{elsarticle}

\usepackage{graphicx}
\usepackage[english]{babel}
\usepackage[version=3]{mhchem}

\journal{Nuclear Physics A}

\begin{document}

    \begin{frontmatter}

        \title{Two-particle photodisintegration of He-4: $\ce{^{4}_{}He}(\gamma,d)d,\ce{^{4}_{}He}(\gamma,p)T,\ce{^{4}_{}He}(\gamma,n)\ce{^{3}_{}He}$}

            \author[khnu]{O.E.~Koshchii}
            \ead{alex.koshchii@gmail.com}

            \author[iert]{P.E.~Kuznietsov}
            \ead{kuznietsov@ukr.net}

            \author[iert]{Yu.A.~Kasatkin}
            \ead{yukasatkin2007@yandex.ru}

            \address[khnu]{V.N.~Karazin Kharkiv National University, 61022, Ukraine, Kharkiv, Svobody Sq.4}
            \address[iert]{Institute of Electrophysics and Radiation Technologies NAS of Ukraine 61002, Ukraine, Kharkiv-2 str. Chernyshevsky, 28, PO Box 8812}

                \begin{abstract}
                    Using a covariant diagram technique and the concept of a nucleus as an elementary particle we calculated differential cross-sections for two-particle photodisintegration reactions of the $\ce{^{4}_{}He}$. The only functional parameter was the vertex structure function, which described the "collapse" of a $\ce{^{4}_{}He}$ nucleus and the nucleon remnants. The interaction of a real photon was determined by the value of particles charge, since the electromagnetic form factors were calculated in the photon point. The inseparability property of the electric charge from the particle mass allowed us to match the energy-momentum and charge conservation laws in the interaction. Therefore, the requirement of gauge symmetry was immediately satisfied. The covariant amplitude of the process equaled to the sum of pole diagrams and regular part\footnote{The regular part fulfills the EM current conservation requirement.}.
                \end{abstract}


\end{frontmatter}

    \section {Introduction}

        Quantum theory of gauge fields (QTGF) is a widely recognized basis of the elementary particle physics. QTGF is based on the thesis that all the known interactions in nature are transferred by means of gauge fields. The principle of gauge symmetry is one of the most important heuristic principles. The considerable success in understanding properties of electromagnetic and weak interactions was achieved by using the gauge invariance principle. This principle implementation led to the modern electroweak interaction theory.

        Quantum electrodynamics (QE) as a important part of QTGF was formed in order to describe electromagnetic (EM) and charged matter fields interactions. The considerable breakthrough in the description of EM interactions was achieved on the basis of the following assumption\footnote{The assumption was assured by the universality property.}: the interaction of a gauge field with fundamental matter fields has a local nature.

        However, the matter fields that form the nature variety are of nonlocal formation and closely related to the strong interaction. Thus, there is an additional uninvestigated interaction in EM processes, which forms a bounded state and at the same time is an origin of all difficulties in using the lagrangian approach. The approach that considers a possibility to construct an analogous Lagrangian that describes a nonlocal and EM field interaction was proposed in the papers \cite{Kasatkin_2004,Kasatkin_2009}. It satisfies the universality principle. The method is based on bringing into consideration the theory of fiber spaces, in which an electromagnetic field vector-potential provides a connection. Due to the fact that the electric charge cannot be separated from the mass and, consequently, it is not an independent quantity, it is necessary to consider additionally the movements in associated charge space, while describing the particles movement in the base space. As the result of these operations it was succeed to harmonize the 4-momentum and charge conservation laws in the amplitude of processes. In the papers \cite{Kasatkin_2009,Kasatkin_2010}, using generalized Feynman rules, the mentioned approach was applied to describe a photodisintegration processes of light nuclei into fragments.

        It is also necessary to follow the requirement of a general covariance and dynamically take into account the requirement of a gauge invariance, while considering the electromagnetic field and nonlocal matter fields interaction. The only and unknown parameter is the vertex that describes the collapse of strongly interacting particles. The dependence of the vertex from the fragments space-like four-momentum allowed us to keep invariance of the approach irrespective of its explicit form.

        Several points that are infusion into the theory by new approach, or, to be more precise, by the regular (pole) part of the amplitude, should be noted.

        Momentum distributions of the components in various nonlocal fields of matter are individual and contain information about the steady-state interactions in a coupled system. They also reflect a spatial and time evolution of a coupled system during the whole energy and structure range. The information for the each nonlocal field is determined by the following things: the decrement of the momentum distribution function, its rate of change and the nature of the curve curvature (its convexity or concavity).

        Another established property of the generalized gauge-invariant pole amplitude, which occurs independently of the explicit form of vertex function, is related to the degree of its increase or decrease. The relative sign between the pole and regular parts in the amplitude is fixed by the total electromagnetic current conservation requirement. If the vertex function of the strong interaction is constant, then the regular part in the amplitude turns into a zero. At the same time the pole part is determined by Yukawa asymptotic behavior - a constant that is divided by the pole. For decreasing functions its derivative is negative. This fact changes the sign in the amplitude for the regular part, making the sign equal to the sign of the pole part. In this case, the contribution from the regular part to the total cross section is constructive (positive interference). In the case, when the vertex function increases with the argument increase, its derivative is positive and the contribution to the cross section is changed to the destructive one.

        To sum up, the regular component of the generalized pole amplitude is a dynamic measure of the nonlocality of a bound state. It shows how "quickly" the structural formations of the initial level of matter structure lose their identity upon transition to the other scale of spatial and temporal localization.

        The amplitude regular component introduces an additional dependence from the vertex function in the form of its derivative. It was established in the paper \cite{Kasatkin_2004} that for the electric dipole splitting contribution from the regular part to the full amplitude at low energies is determined by the derivative from the strong interaction vertex. If the electric dipole transition is absent (the case of splitting into two identical fragments) than a regular part contribution to the total amplitude is determined by the second derivative from the strong interaction vertex.

        Investigation of interaction processes of the EM field with nuclei appears to be the important method. It helps to solve the vast number of nuclei and elementary particle physics issues. Namely: an understanding of the role of different reaction mechanisms, a revision of different nuclei models, an obtaining of information about nucleon-nucleon interaction, an analysis of the structure of nuclei wave functions, an understanding of the role of quark configurations and non nucleon degrees of freedom in nuclei, etc. In the theoretical aspect, this method possesses the significant advantage -- we can consider processes with the bounds of perturbation theory using the constant of the EM interaction.

        Especially valuable results can be obtained while investigate EM interactions of few-nucleon systems. The special place in nuclear physics belong to these systems for the number of reasons. Some of them are the relative simplicity of structure of such nuclei and the possibility to find the precise solution for the tree- or four-body problem. Thus, applying the proposed theory we considered two-particle photodisintegration processes of He-4 in our paper. A close fit of theoretical calculations and experimental measurements on differential and total cross sections for these reactions was obtained by using a minimum number of parameters.

    \section {Process $\ce{^{4}_{}He}(\gamma,d)d$}

        The process $\ce{^{4}_{}He}(\gamma,d)d$ is characterized by the fact that due to isospin selection and identity of particles in a final state the electric dipole moment is suppressed and the process realizes mostly due to quadrupole $\gamma$-ray absorption. Therefore, this channel is of a considerable interest to study the nature of a quadrupole transition.

        For the compact cross-section notation of the process $\ce{^{4}_{}He}(\gamma,d)d$ it is convenient to use spiral amplitudes, which were obtained in Cartesian coordinate system when the $\gamma$-quantum momentum is directed along $z$-axe and the first deuteron momenta is situated in the $xOz$ plane. Initially, we defined the following matrix:

        \begin{equation}\label{eq:1}
            {R_{{\lambda _\gamma },\lambda _\gamma'}} = \sum\limits_{\{ \lambda \} } {M_{{\lambda _1},{\lambda _2}}^{{\lambda _\gamma }}{\rho _{{\lambda _\gamma },\lambda _\gamma'}}M_{{\lambda _1},{\lambda _2}}^{ * \lambda _\gamma'}}
        \end{equation}
        where ${M_{{\lambda _1},{\lambda _2}}^{{\lambda _\gamma }}}$ are spiral amplitudes of the $\ce{^{4}_{}He}(\gamma,d)d$ process, ${\lambda _\gamma },{\lambda _1}$ and ${\lambda _2}$ are helicities of a $\gamma$-quantum, first and second deuterons respectively, and ${\rho _{{\lambda _\gamma },\lambda _\gamma'}}$ is a polarization density matrix of a $\gamma$-quantum.

\begin{figure}[htp]
            \centering
            \includegraphics{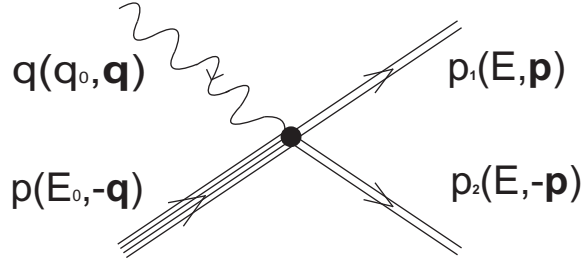}
            \caption{$\ce{^{4}_{}He}(\gamma,d)d$ process in the center-of-mass system.}
            \label{fig:1}
        \end{figure}

        The differential cross section of the $\ce{^{4}_{}He}(\gamma,d)d$ process in the case when a $\gamma$-ray is polarized in an arbitrary way in the center-of-mass system (see the Fig.\ref{fig:1}) is:

        \begin{equation}\label{eq:2}
            \frac{{\partial \sigma }}{{\partial \Omega }} = \frac{1}{{2{\left( {8\pi W} \right)}^{2}}}\frac{{\left| \bf p \right|}}{{\left| \bf q \right|}}SpR,
        \end{equation}
        where:
        \[W = {q_{_0}} + {E_0} = 2E.\]

        Using the Eq.\ref{eq:1} and the explicit form of a photon polarization density matrix we obtained:

        \begin{eqnarray*}
            SpR = \sum\limits_{{\lambda _1},{\lambda _2}} {M_{{\lambda _1},{\lambda _2}}^1} M_{{\lambda _1},{\lambda _2}}^{1*} + 2{\mathop{\rm Re}\nolimits} ({M_{1,1}}M_{ - 1, - 1}^* + {M_{1, - 1}}M_{ - 1,1}^* - \\
            - {M_{1,0}}M_{ - 1,0}^* - {M_{0,1}}M_{0, - 1}^* + \frac{1}{2}{\left| {{M_{0,0}}} \right|^2})
        \end{eqnarray*}

\begin{figure}[htp]
            \centering
            \includegraphics{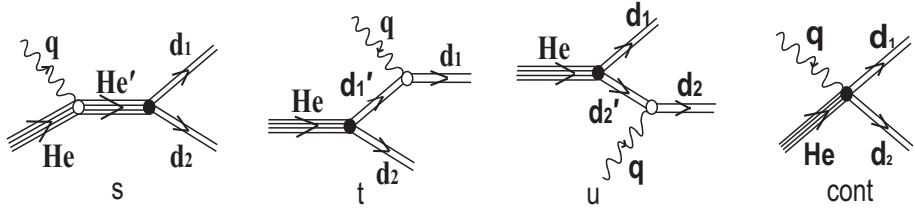}
            \caption{Set of diagrams for the $\ce{^{4}_{}He}(\gamma,d)d$ reaction.}
            \label{fig:2}
        \end{figure}

        In order to write down the amplitude of the process $\ce{^{4}_{}He}(\gamma,d)d$ we followed the approach described in the paper \cite{Kasatkin_2009}. In this paper the problem of ensuring a gradient invariance of the amplitude was solved by choosing the next reaction mechanism: the contact diagram was added to the known field-theoretic row. This diagram takes into account multiparticle effects, including the electromagnetic interaction with the "carriers of strong interaction". Rooting from this approach, the amplitude, which satisfies the principles of relativistic and gradient invariance, was determined by the sum of a contact and pole diagrams, which are shown on the Fig.\ref{fig:2}.

\begin{figure}[htp]
            \centering
            \includegraphics[scale=0.38]{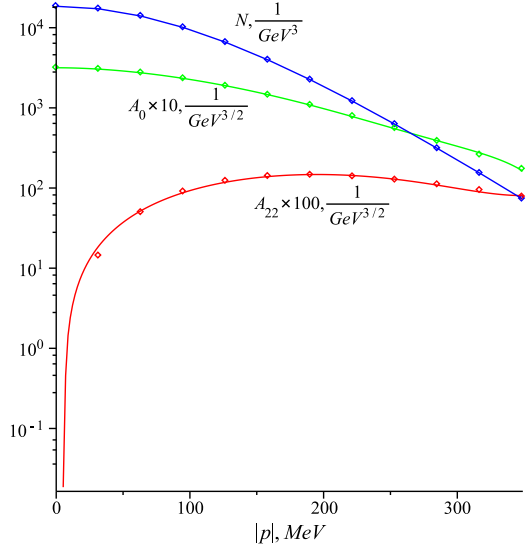}
            \caption{S-, D-waves and momentum distribution approximation for the Urbana function.}
            \label{fig:3}
        \end{figure}

        The matrix element pole part of these diagrams is characterized by the two electromagnetic vertices $(\gamma \ce{^{}^4He} \to \ce{^{}^4He})$, $(\gamma d \to d)$ and the strong one - $\ce({}^4He \to dd)$. The explicit form of these vertices can be obtained by the use of Argonne \cite{Zayats_1994} or Urbana parametrization. To obtain these vertices using the Urbana parametrization we used data from the \cite{Schiavilla_1986}. The following parametrization of functions $A_{dd}^{00}$ and $A_{dd}^{22}$ is possible:
        \[A_{dd}^{00}(\left| \bf p \right|) = a1 + a2*\left| \bf p \right| + a3*{\left| \bf p \right|^2} + a4*{\left| \bf p \right|^3} + a5*{\left| \bf p \right|^4},\]
        \[A_{dd}^{22}(\left| \bf p \right|) = b1 + b2*\left| \bf p \right| + b3*{\left| \bf p \right|^2} + b4*{\left| \bf p \right|^3} + b5*{\left| \bf p \right|^4},\]
        where parameters $a1 = 316.513,a2 = 0.484*{10^{ - 3}},a3 =  - 0.0138,a4 = 0.543*{10^{ - 4}},a5 =  - 6.21*{10^{- 8}}$;$b1 =  - 0.0272,b2 = 0.373*{10^{ - 2}},b3 = 0.117*{10^{- 3}},b4 =  - 0.6847*{10^{- 6}},b5 = 9.696*{10^{- 10}}$. The Fig.\ref{fig:3} shows the functions $A_{dd}^{00}(\left| \bf p \right|)$, $A_{dd}^{22}(\left| \bf p \right|)$ and $N_{dd}(\left| \bf p \right|)$ dependence, where solid lines represent the approximation of real values that marked as dots.

        Taking into account the expressions for vertices \cite{Zayats_1994} we estimated the following matrix elements corresponding to the pole s-, t- and u-channel diagrams:

        \[M^{(s)}  = e\varepsilon _\mu  (2p + q)^\mu  \frac{1}{{s - m_{He}^2 }}U_\rho ^* (p_1 )U^{*\sigma } (p_2 )G_\sigma ^\rho  (p;p_1 ,p_2 ),\]
        \[M^{(t)}  = e_1 \varepsilon _\mu  F_\beta ^{\mu \rho } (q,p'_1 ,p_1 )\frac{1}{{t - m_d^2 }}U_\rho ^* (p_1 )U^{*\sigma } (p_2 )G_\sigma ^\beta  (p;p'_1 ,p_2 ),\]
        \[M^{(u)}  = e_2 \varepsilon _\mu  F_\sigma ^{\mu \beta } (q,p'_2 ,p_2 )\frac{1}{{u - m_d^2 }}U_\rho ^* (p_1 )U^{*\sigma } (p_2 )G_\beta ^\rho  (p;p_1 ,p'_2 ),\]
        where $s = {(q + p)^2},t = {({p_1} - q)^2},u = {({p_2} - q)^2}$ are Mandelstam variables, ${m_d}$ and ${m_{He}}$ are a deuteron and a helium nucleus masses respectively, $p' = p + q$.

        The matrix element which corresponds to the contact diagram is presented in the integral form:

        \[{M^{(c)}} = e{\varepsilon _\mu }U_\rho ^*({p_1})U^{*\sigma } (p_2 ) \times \]
        \[ \times \int\limits_0^1 {\frac{{d\lambda }}{\lambda }} \frac{\partial }{{\partial {q_\mu }}}\left[ {{e_1}{G_\sigma ^\rho}(p' - q\lambda ;{p_1} - q\lambda ,{p_2}) + {e_2}{G_\sigma ^\rho}(p' - q\lambda ;{p_1},{p_2} - q\lambda )} \right],\]

        In the above-mentioned model the full amplitude of the process $\ce{^{4}_{}He}(\gamma,d)d$ is determined by the following sum: $M^{(s)}  + M^{(t)}  + M^{(u)}  + M^{(c)}$. A differential and a total cross-sections were calculated by substitution the full amplitude into the Eq.\ref{eq:2}.

\begin{figure}[htp]
            \centering
            \includegraphics[width=4in,height=3in]{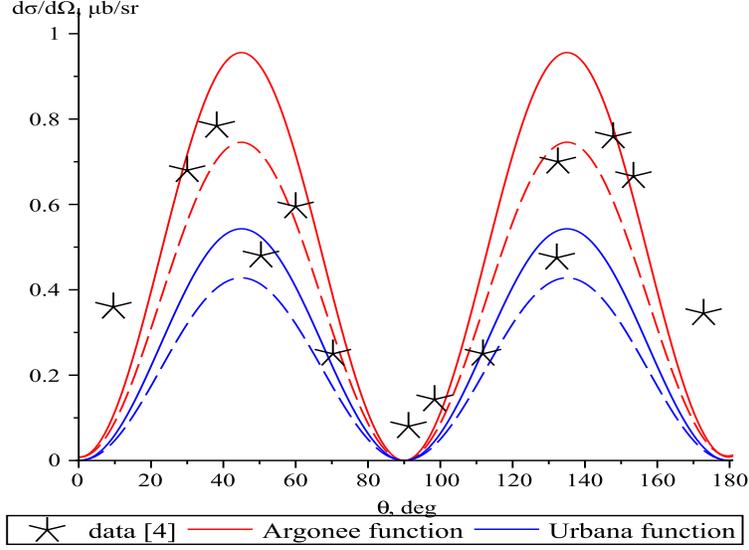}
            \caption{$\ce{^{4}_{}He}(\gamma,d)d$ differential cross section at $E_\gamma=40$ MeV, solid lines - include contact part, dash lines - without it.}
            \label{fig:4}
        \end{figure}

        The Fig.\ref{fig:4} shows a differential cross section angular dependence of the process $\ce{^{4}_{}He}(\gamma,d)d$ at photon energies in lab system $E_\gamma=40$ MeV for the case when a $\gamma$-quantum is linearly polarized. A qualitative description of the experimental angular distribution was obtained: the correct location of the cross-section minimum at $\upsilon  = 90^ \circ$ and maximums at $\upsilon  = 45^ \circ,135^ \circ$.

        Quadrupole transition can be investigated by the analysis of the differential cross-section at angles $\upsilon  = 0^ \circ,90^\circ$ and $180^\circ$. To thoroughly investigate the nature of the dipole transition we should proceed to the next section.

    \section{$\ce{^{4}_{}He}(\gamma,p)T$ and $\ce{^{4}_{}He}(\gamma,n)\ce{^{3}_{}He}$ reactions}

        To estimate the differential cross-section of the processes $\ce{^{4}_{}He}(\gamma,N)T$, where $N$ is  a nucleon(either $p$ or $n$) and $T$ is either $\ce{^{3}_{}He}$ or $\ce{^{3}_{}H}$, we have to write down the corresponding matrix element. It equals:

        \[M = e{\varepsilon _\mu }\bar u(N)\sum\limits_{i = s,t,u,c} {{M^{\mu (i)}}\nu (T),\nu (T) = C{{\bar u}^T}(T)} \]
        where:
        \[{M^{\mu (s)}} = e\frac{{{{(p + p')}^\mu }}}{{s - m_{He}^2}}{G^{(s)}}{\gamma _5},\]
        \[{M^{\mu (t)}} = {j^{\mu (t)}}\frac{{(\hat N' + {m_N})}}{{t - m_N^2}}{G^{(t)}}{\gamma _5},\]
        \[{M^{\mu (u)}} = {G^{(u)}}{\gamma _5}\frac{{(\hat T' - {m_T})}}{{u - m_T^2}}{j^{\mu (u)}},\]
        \[{M^{\mu (c)}} = \int\limits_0^1 {\frac{{d\lambda }}{\lambda }\frac{\partial }{{\partial {q_\mu }}}\{ {z_N}G[ - k_{st}^2(\lambda )] + {z_T}G[ - k_{su}^2(\lambda )]\} } {\gamma _5},\]
        $q,p,N$ and $T$ are 4-momenta of a $\gamma$-quantum, $\ce{^{4}_{}He}$, nucleon and a nucleus T respectively. Electromagnetic currents were defined in a standard way: ${j^{\mu (t)}} = ({z_N} + {k_N}\hat k){\gamma ^\mu },{j^{\mu (u)}} = ({z_T} + {k_T}\hat k){\gamma ^\mu }$, where ${z_N}({z_T})$ and ${k_N}({k_T})$ are the charge and anomalous magnetic moment of a particle $N(T)$; ${z_H}$ is a charge of a $\ce{^{4}_{}He}$.

        Relative four-momenta that characterize the vertex $\ce{^{4}_{}He} \to NT$ in the pole diagrams equals:
        \[{k_s} = N - \frac{{(Np')}}{{{{p'}^2}}}p' = \frac{{(Tp')}}{{{{p'}^2}}}p' - T,\]
        \[{k_t} = {k_s} - \frac{{(Tp')}}{{{{p'}^2}}}q,{k_u} = {k_s} + \frac{{(Np')}}{{{{p'}^2}}}q.\]

        Quantities $k_{st}(\lambda)$ and $k_{su}(\lambda)$ were defined as:
        \[{k_{st}}(\lambda ) = {k_s} - \lambda \frac{{(Tp')}}{{{{p'}^2}}}q,{k_{su}}(\lambda ) = {k_s} + \lambda \frac{{(Np')}}{{{{p'}^2}}}q.\]

        Vertex functions ${G^{(i)}} \equiv G( - k_i^2),(i = s,t,u)$ depend on the appropriate four-momentum. They describe the virtual collapse of a $\ce{^{4}_{}He}$ into $NT$ and, due to relativistic invariance depend on the square of a relative four-momentum.

        It is also should be noted that in the case when ${G^{(s)}} = {G^{(t)}} = {G^{(u)}} = const$ we have ${M^{\mu (c)}}=0$. Therefore, the sum of pole diagrams is a gauge-invariant quantity.

\begin{figure}[htp]
            \centering
            \includegraphics[scale=0.36]{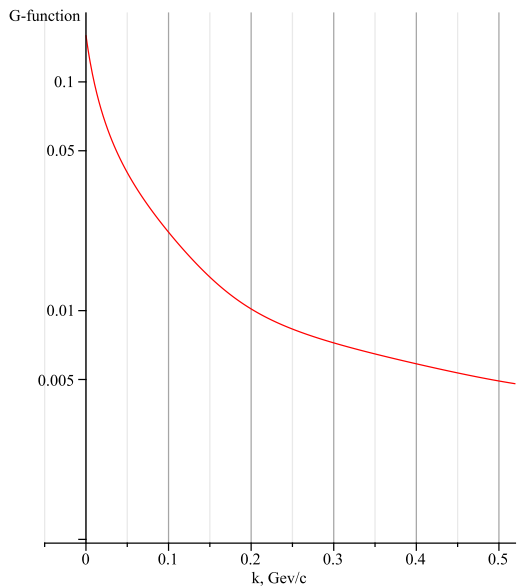}
            \caption{Energy dependance of a $G$-function.}
            \label{fig:5}
        \end{figure}

        We parameterized our vertex function on the basis of the paper \cite{Schiavilla_1986}. This step allowed us to determine all necessary quantities. Fig.\ref{fig:5} shows the parametrization of a $G$ as a function of the fragments relative momentum.

\begin{figure}[htp]
            \centering
            \includegraphics[scale=0.55,angle=0.]{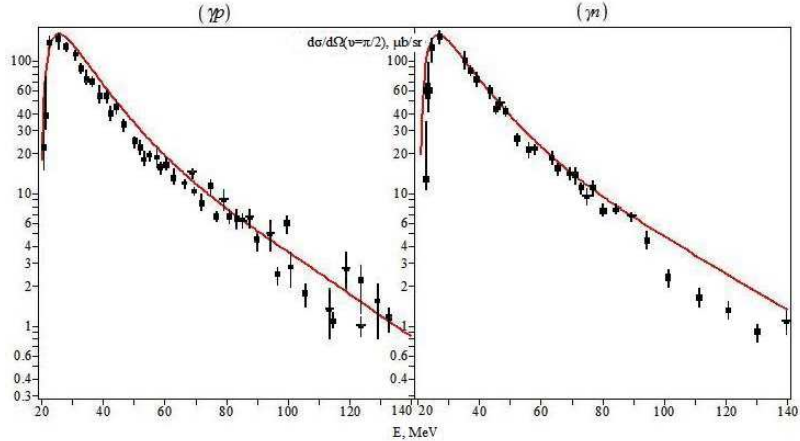}
            \caption{A dependence of the differential cross section $\ce{^{4}_{}He}(\gamma,N)T$  on a photon energy at angles $\upsilon  = {90^ \circ }$. Experimental data from \cite{Gari_1981} and \cite{Klepikov_2009}.}
            \label{fig:6}
        \end{figure}

        As soon as all quantities were defined we made calculations and compared them with experimental data without changing any previously fixed variables. The Fig.\ref{fig:6} shows the dependence of the differential cross-section $\ce{^{4}_{}He}(\gamma,N)T$ on the photon energy at the angle $\upsilon  = {90^ \circ }$($E$ is a photon energy in a laboratory system). The obtained data fits well with experimental ones. According to the results, a standard pole amplitudes row should be supplemented with an additional mechanism, namely regular part of the amplitude, to better describe considering processes. Practically at all photon energies the regular part is essential.

\begin{figure}
            \centering
            \includegraphics[scale=0.38]{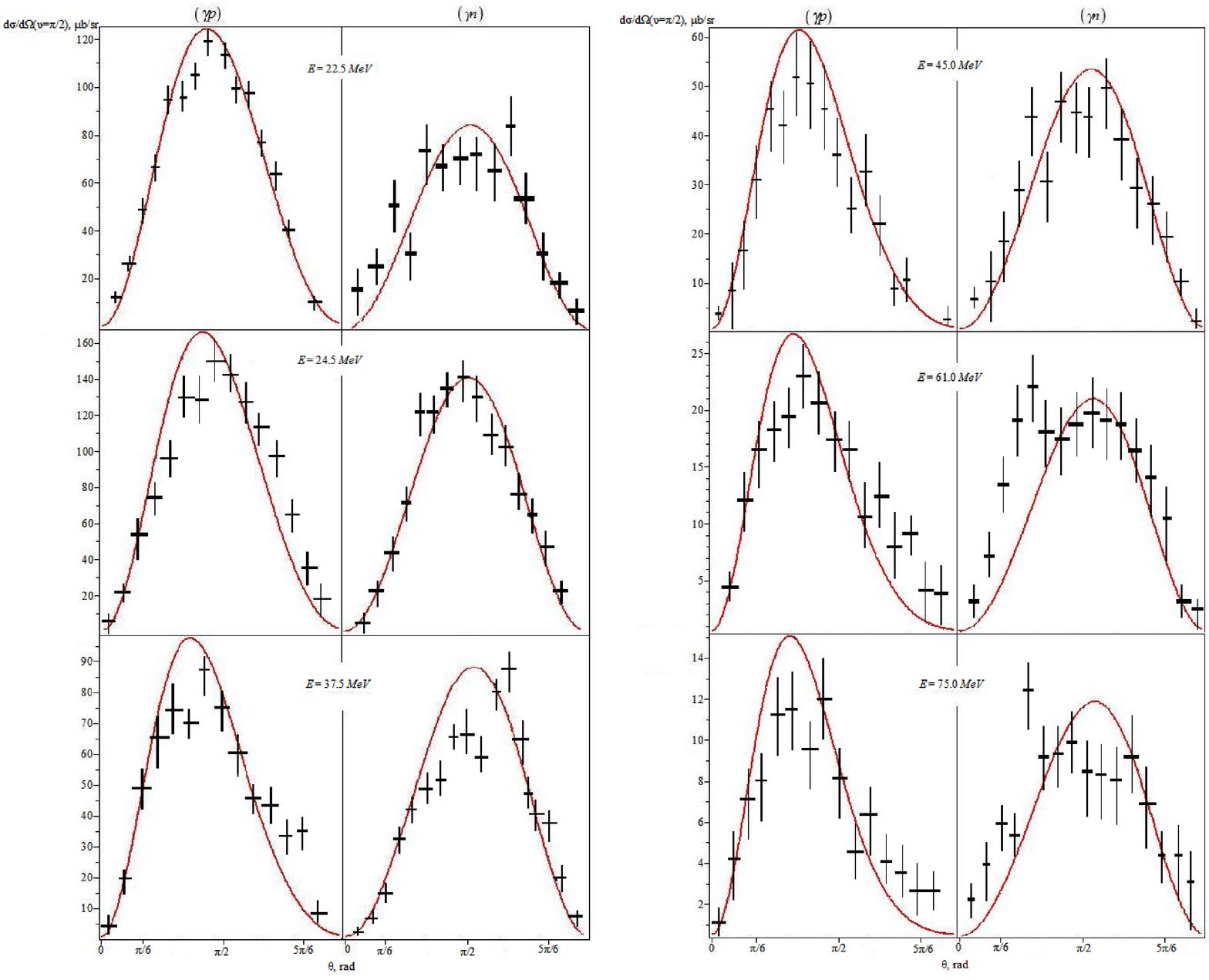}
            \caption{An angular dependence of differential cross sections for reactions $\ce{^{4}_{}He}(\gamma,N)T$ at the energy interval from  ${E_\gamma }=22.5$  to $75.0$ MeV. $\dagger$ - data from \cite{Klepikov_2009}.}
            \label{fig:7}
        \end{figure}

        The Fig.\ref{fig:7} presents six pairs of the angular spectra at fixed energies. The experimental angular distributions are well described by this model at any considered energy interval.

\begin{center}%
        \begin{figure}
            \includegraphics[scale=0.5,angle=270.]{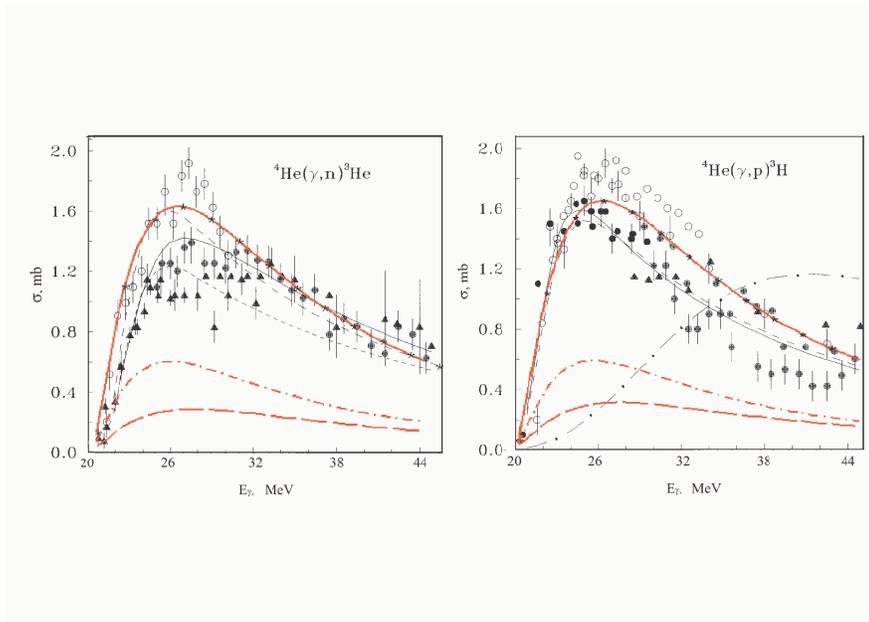}
            \caption{A dependence of total cross sections for reactions $\ce{^{4}_{}He}(\gamma,N)T$ on a photon energy at the the interval from 20 to 44 MeV. Experimental data from \cite{Dubovichenko_2004}.}
            \label{fig:8}
        \end{figure}
        \end{center}

        In the Fig.\ref{fig:8} the red solid line indicates dependence of total cross sections for reactions $\ce{^{4}_{}He}(\gamma,N)T$ on the photon energy interval from 20 to 44 MeV taking into consideration all the diagrams. Dash-dotted and dotted lines describe the accounting of pole diagrams and the regular one, respectively. It is evident that the required agreement with experimental data can be achieved only when we consider both of inputs. Accounting only the pole diagrams does not provide an adequate description of experimental data.

    \section {Conclusion}

        We attained the high-quality theoretical description of two-particle photodisintegration processes of He-4, using minimal number of parameters. The fulfilled calculations and comparisons with experimental measurements have shown that a generalization of the Feynman rules for the description of photonuclear processes works. The alternative approach to the theory describing electromagnetic processes of compound systems allowed us to reproduce the results without any problems. The special role was given to the construction of a regular part of the amplitude, which determined the gauge-closed matrix element. It means that the structure of the matrix element has been adapted to the description of various processes. These elements satisfied the requirements of covariance and the fundamental requirement of gauge symmetry.

    \section*{Acknowledgement}
        We would like to express our sincere gratitude to Dr. Ratkevich S.S. and Mr. Dubovoy M. for their constructive suggestions and useful pieces of advice.


\end{document}